\documentclass[
 reprint,
 runinaddress,
 amsmath,amssymb,
 aps,
]{revtex4-2}
\usepackage{graphicx}
\usepackage{dcolumn}
\usepackage{bm}
\usepackage[separate-uncertainty = true]{siunitx}
\DeclareSIUnit{\molar}{M}

\begin{document}

\preprint{APS/123-QED}

\title{Near-perfect efficiency in X-ray phase
microtomography}
\author{Dominik John$^{1,2,*}$, Gregor Breitenhuber$^{1}$, Sami Wirtensohn$^{1,2}$, Franziska Hinterdobler$^{1}$,\\ Luka Gaetani$^{1}$, Sara Savatovi{\'c}$^{1}$, Jens Lucht$^{3}$, Markus Osterhoff$^{3}$,\\ Marina Eckermann$^{4}$, Tim Salditt$^{3}$, and Julia Herzen$^{1}$}
\affiliation{\parbox{0.8\textwidth}{%
$^{1}$Research Group Biomedical Imaging Physics, Department of Physics,\\ TUM School of Natural Sciences \& Munich Institute of Biomedical Engineering,\\Technical University of Munich, James-Franck-Straße 1, 85748 Garching, Germany\\
$^{2}$Helmholtz-Zentrum Hereon, Institute of Materials Physics, 21502 Geesthacht, Germany\\
$^{3}$Institut für Röntgenphysik, Georg-August-Universität Göttingen,\\ \phantom{$^{3}$}Friedrich-Hund-Platz 1, 37077 Göttingen, Germany\\
$^{4}$Institute of Applied Physics, Universität Bern, Sidlerstrasse 5, 3012 Bern, Switzerland\\
$^{*}$Corresponding author: dominik.john@tum.de
}}

\begin{abstract}
X-ray microtomography at synchrotron sources is fundamentally limited by the high radiation dose applied to the samples, which restricts investigations to non-native tissue states and thereby compromises the biological relevance of the resulting data. The limitation stems from inefficient indirect detection schemes that require prolonged exposures. Efforts to extract additional contrast through multimodal techniques, like modulation-based imaging, worsen the problem by requiring multiple tomographic scans. In addition, the techniques suffer from low modulator pattern visibility, which reduces measurement efficiency and sensitivity.
We address both the detection efficiency and modulation visibility challenges using a novel setup that combines an X-ray waveguide, a structured phase modulator, and a photon-counting detector. Our approach simultaneously achieves near-theoretical limits in both visibility (\SI{95}{\percent}) and quantum efficiency (\SI{98}{\percent}), thereby enabling dose-efficient multimodal microtomography at single-micrometer resolution. This advance will enable new classes of experiments on native-state biological specimens with the potential to advance biomedical research, disease diagnostics, and our understanding of tissue structure in physiological environments.
\end{abstract}

\maketitle

\section{Introduction}
X-ray micro-computed tomography (µCT) has enabled valuable three-dimensional insights into biological specimens \cite{thiesse2010lung, schambach2010application, eckermann20203d, reichardt20213d, shakeri2019can, metscher2009microct1, metscher2009microct2, friedrich2008micro} and is often considered a non-destructive imaging technique. However, recent investigations have revealed that radiation damage at synchrotron sources poses a significant challenge: Intense X-ray exposure causes severe cellular damage \cite{petruzzellis2018pitfalls}, with structural changes detectable from the first moments of exposure \cite{sauer2022primary}. These pitfalls limit the reliability of imaging-derived data in biological research \cite{petruzzellis2018pitfalls}. High photon flux also produces radicals in wet specimens~\cite{fornasiero2014super}, leading to gas bubble formation and sample deformation. To prevent this, samples are typically embedded in paraffin \cite{riedel2023comparing} or ethanol~\cite{john2024centimeter}, neither of which preserves the native tissue environment.

While computational approaches are being investigated to mitigate dose requirements \cite{obata2025enhancing}, the root of the dose problem lies in the detection technology itself. High-resolution µCT setups at synchrotrons and laboratories employ CCD or CMOS chips coupled to scintillators \cite{longo2024syrmep}, which present a fundamental trade-off between spatial resolution and efficiency: Thicker scintillators increase efficiency but introduce signal blurring. The poor quantum efficiency of these indirect detectors means that much of the radiation passing through the sample is not detected, necessitating higher doses to achieve acceptable image quality. Photon-counting (or direct) detectors circumvent this trade-off, providing dose-efficient imaging through high quantum efficiency~\cite{sellerer2019quantitative, gkoumas2019dose}, noise-free readout, and sharp pixel response across high dynamic ranges. These advantages have been successfully demonstrated at laboratory sources~\cite{sellerer2019quantitative, scholz2020biomedical}. However, the comparatively large pixel size of at least \SI{55}{\micro\meter}~\cite{pennicard2018lambda} has precluded the use of direct detectors for high-resolution synchrotron microtomography, where effective pixel sizes in the single-micrometer range are required.

The challenge of dose-efficient imaging is further complicated by the need for enhanced contrast. Many biological tissues exhibit low contrast in conventional attenuation-based imaging, necessitating the development of methods to additionally exploit the phase shift and scattering introduced by the samples. Some of these methods rely on the free-space propagation of the X-ray wavefront as a contrast mechanism and include propagation-based imaging~\cite{snigirev1995possibilities, cloetens1996phase, paganin2002simultaneous}, holography~\cite{cloetens1999holotomography}, and ptychography~\cite{pfeiffer2018x}. Other techniques employ additional optical elements such as crystal analyzers~\cite{bravin2003exploiting}, gratings~\cite{weitkamp2005x, bennett2010grating}, or various wavefront markers~\cite{morgan2012x, berujon2012two, gustschin2021high}. Speckle-based imaging~\cite{morgan2012x, zdora2017x, zdora2018state}, also more generally referred to as modulation-based imaging to include structured diffusers~\cite{quenot2022x, savatovic2025high}, uses wavefront markers to track changes and displacements in the X-ray wavefront imparted by a sample. This is performed either by explicitly tracking changes in subregions of the image~\cite{berujon2016x, zdora2017x} or using a global algorithmic approach~\cite{quenot2021implicit, alloo2023m}. One advantage of the modulation-based approach compared to optics-free methods is that it enables the simultaneous extraction of three imaging modalities related to the sample's properties: transmission, phase, and small-angle scattering (also known as dark field). In addition, the method enables a quantitative rather than qualitative access to the phase of the examined object in the form of electron density maps~\cite{zdora2020x} free of prior assumptions on the sample composition~\cite{riedel2023comparing}. However, these benefits come at the cost of acquiring multiple tomographic scans at different modulator positions, leading to increased measurement time and sample dose. 

The efficiency of modulation-based imaging critically depends on achieving high visibility, which characterizes the intensity differences in the modulation pattern and directly relates to signal-to-noise ratio~\cite{yan2020predicting, neuwirth2020high}. High visibility enables analysis of weaker signals and permits shortened measurement times, making it essential for measurement and dose efficiency. To address this need, 2D Talbot array illuminators (TAIs) have been recently introduced to increase visibility~\cite{gustschin2021high}. Unlike random modulators, such as sandpaper, these optical elements create phase shifts in the wavefront in a regular pattern, with the shift magnitude optimized for maximum visibility at specific energies and measurement distances. While advances in the modulator provide one avenue for increasing visibility, increasing the coherence of the source provides another complementary approach. X-ray waveguides have emerged as a useful tool for creating highly coherent point sources~\cite{pfeiffer2002two}. Combined with suitable focusing optics~\cite{jarre2005two, salditt2015compound}, waveguide-based sources have, for example, been successfully applied to investigate human pathology~\cite{eckermann20203d, reichardt20213d} using propagation-based imaging.

To simultaneously overcome the mentioned dose efficiency and visibility limitations of conventional micro-computed tomography setups, we propose a novel setup that combines X-ray waveguides with Talbot array illuminators and a photon-counting detector, which has a quantum efficiency of \SI{98}{\percent} at \SI{8}{\kilo\electronvolt}~\cite{dectris2019eiger}. We experimentally characterize the setup's properties by measuring the achieved visibilities over a range of propagation distances and determine the angular sensitivity using both a direct and an indirect detector. Additionally, we perform and analyze a phase tomography of a biological sample.

\section{Results}
\subsection{Setup Design}
\begin{figure*}[htb]
    \centering    \includegraphics[width=.8\linewidth]{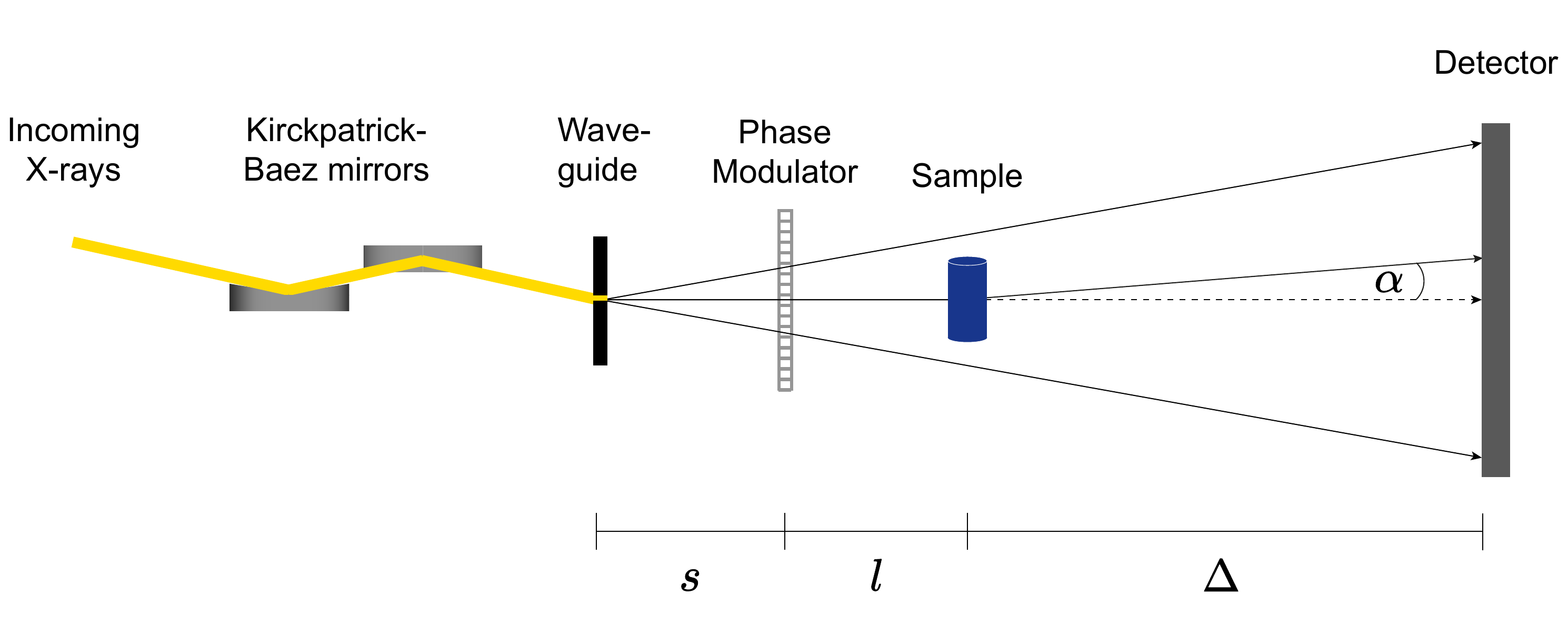}
    \caption{The radiation produced by an undulator source and filtered by a monochromator (not shown) is focused onto a waveguide via a Kirkpatrick-Baez mirror. A phase modulator is placed a distance $s$ downstream of the waveguide, and the sample is located a distance $l$ downstream of the Talbot array. The detector is placed a distance $\Delta$ downstream of the sample. Due to the presence of the sample, phase effects lead to a refraction of the reference pattern by an angle $\alpha$.}
    \label{fig:setup}
\end{figure*}
A schematic drawing of the proposed setup is shown in Fig.~\ref{fig:setup}. The main components are the Kirkpatrick-Baez mirrors, which focus the incoming X-rays onto an X-ray waveguide. The waveguide allows for a small focal spot and a smooth wavefront. A phase modulator in the form of a Talbot array illuminator is placed a distance $s$ downstream of the waveguide. Both a photon-counting and a scintillator-based detector were used and compared. More details on the components can be found in Sec.~\ref{sec:experimental-setup}.

\subsection{Visibility}
\label{sec:visibility-results}
\begin{figure*}[htb]
    \centering
    \includegraphics[width=.9\linewidth]{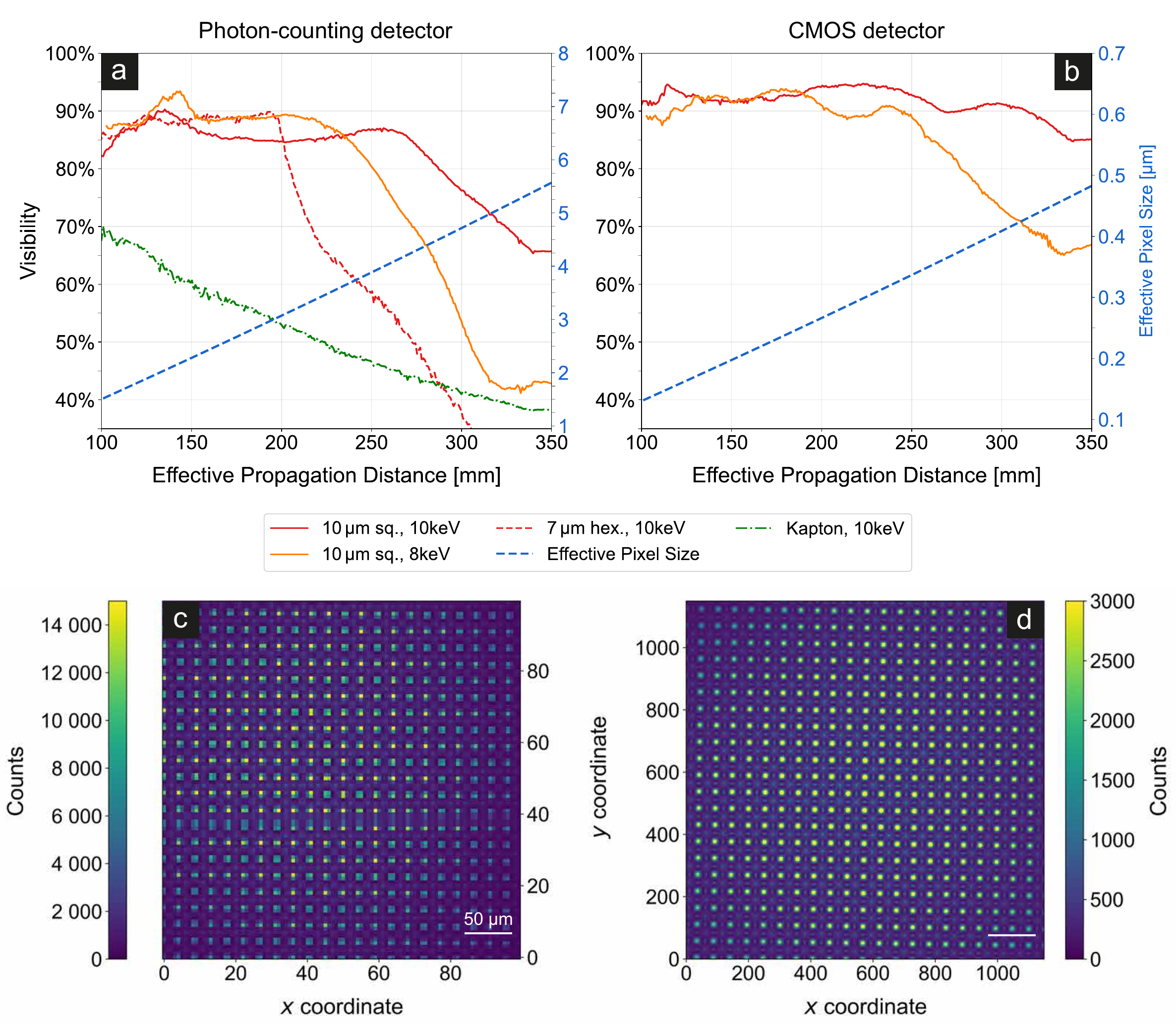}
    \caption{Visibility vs. effective propagation distance for different phase modulators. The top panels show the visibility as a function of the effective propagation distance for both detector types. The photon-counting detector camera (a) achieves a maximum visibility of \SI{93.4}{\percent}, and the CMOS detector (b) reaches \SI{94.8}{\percent}. The blue dashed line indicates how the effective pixel size increases with propagation distance. The bottom panels display the modulation patterns captured by each detector system at their respective maximum visibilities. Both use the \SI{10}{\micro\meter} period TAI. The intensity pattern of the photon counter (c) was captured at a beam energy of \SI{8}{\kilo\electronvolt}, and the pattern of the CMOS detector (d) was imaged at a beam energy of \SI{10}{\kilo\electronvolt}. The strong and regular intensity changes are clearly visible; however, the CMOS detector resolves the pattern better due to its smaller pixel size. The blur at approximately $y = 40$ in the photon-counting detector panel is due to interpolation of the detector gap.}
    \label{fig:visibility-plot}
\end{figure*}

We placed different phase modulators at varying distances downstream of the waveguide, keeping the detector position fixed, and recorded the resulting visibilities. The experiment was conducted at two different beam energies of \SI{8}{\kilo\eV} and \SI{10}{\kilo\eV}. Figure~\ref{fig:visibility-plot} shows a comparison of visibilities for the different modulators for the two detector types: the photon counter and the CMOS detector.

Peak visibility was similar for all structured modulators and was significantly higher than that of the Kapton random diffuser. The CMOS camera achieved a maximum visibility of \SI{94.8}{\percent} and the photon-counting detector reached \SI{93.4}{\percent} for the \SI{8}{\kilo\eV} case using the \SI{10}{\micro\meter} period modulator. As expected, given the fundamental optical properties of the setup, the performance between the two detector types was similar. A key finding is that visibility remains stable over a large range of propagation distances, particularly for the CMOS detector. For the photon-counting detector, the visibility drops as the effective pixel size approaches the modulator period. This is expected, since the modulation pattern is no longer adequately sampled. In contrast, the effective pixel size of the CMOS detector remains small enough to adequately sample the modulation pattern across the entire measurement range.

The visual impression of the recorded modulation patterns confirms the quantitative visibility measurements. Bright intensity maxima are positioned next to regions of almost no intensity in both detector cases, demonstrating the high contrast.

\begin{figure}[htb]
    \centering \includegraphics[width=1.0\linewidth]{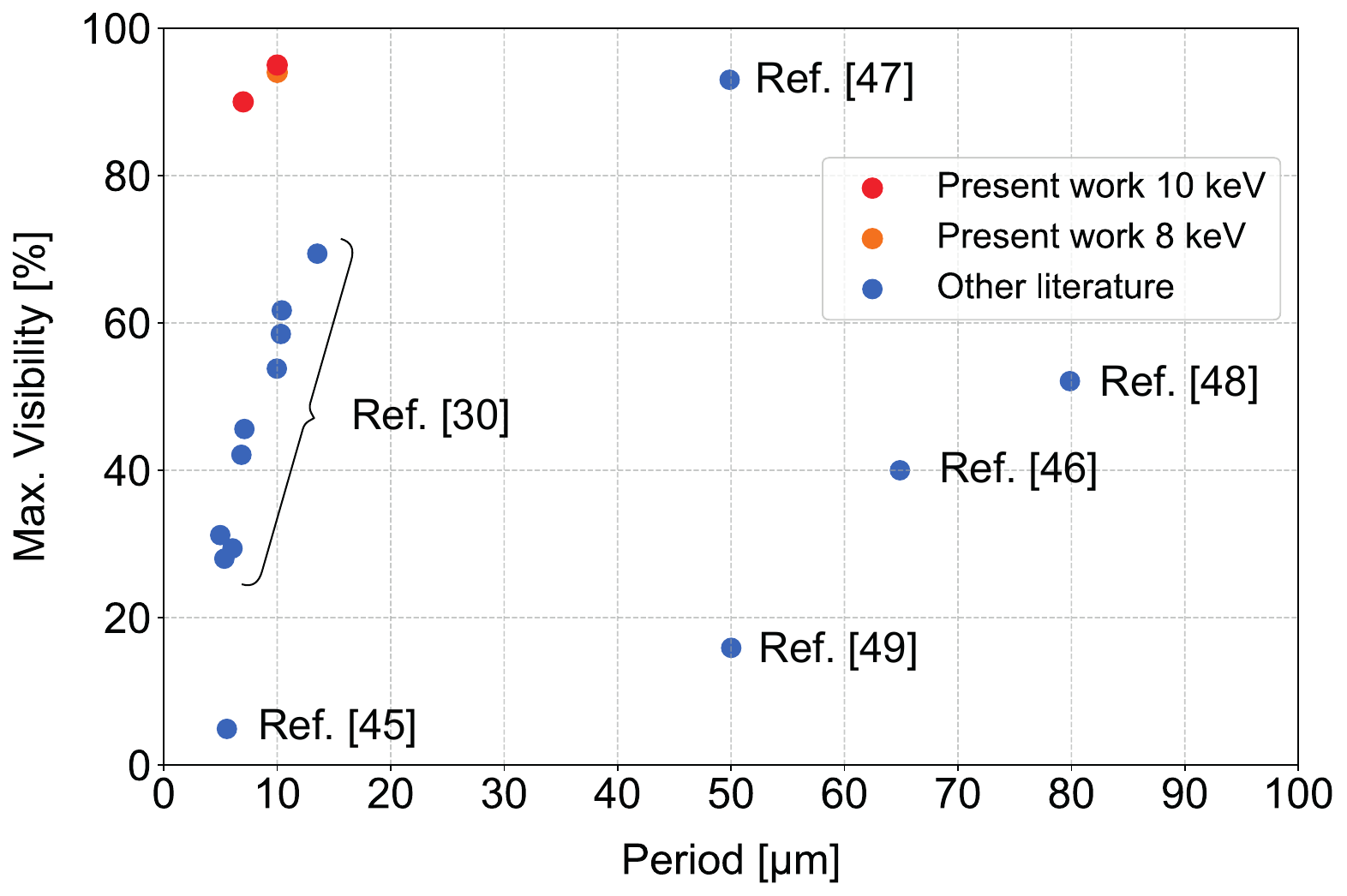}
    \caption{Combinations of visibility and modulator period from previous literature (blue)~\cite{gustschin2021high, morgan2013sensitive, reich2018scalable, dos2018shack, kagias2019diffractive, zakharova2019inverted} and the current work (orange and red). The highest available value for each combination was used, regardless of the detector type. Due to the strong intensity gradients created, a combination of small period size and high visibility is preferable for modulation-based retrieval of phase shifts. Compared to previous literature, our approach enables visibility values close to the theoretical upper limit of \SI{100}{\percent} at small period sizes.}
    \label{fig:period-visibility-comparison}
\end{figure}

Fig.~\ref{fig:period-visibility-comparison} shows different combinations of structure period sizes and the visibility that has been achieved. For imaging applications, smaller period sizes at high visibility are preferable, since they create strong intensity gradients even at the scale of smaller sample features. Although a visibility of $\SI{93}{\percent}$ has been reported for a period of $\SI{50}{\micro \meter}$~\cite{dos2018shack}, a similar value has not yet been achieved for significantly smaller structure sizes. Our reported setup provides a visibility close to the theoretical maximum for period sizes of \SI{10}{\micro\meter} and \SI{7}{\micro\meter}. Moreover, the non-dispersive nature of the compound optics employed, the Kirkpatrick-Baez mirrors and the X-ray waveguide, allows for rapid changes in energy. This was utilized for the switch from \SI{8}{\kilo\eV} to \SI{10}{\kilo\eV}.

\subsection{Angular sensitivity}
The phase sensitivity was determined by acquiring a projection of a test object with 26 modulator positions. The same object was imaged with an exposure time of \SI{1.0}{\second} and a sample distance $s = \SI{260}{\milli\meter}$ using both the photon-counting detector and the CMOS camera. To match the resolution of the photon-counting detector, the CMOS camera data was binned by a factor of 11 prior to phase retrieval. In both cases, the Unified Modulated Pattern Analysis (UMPA) was used to track the local displacement of the pattern, corresponding to the refraction angles. The UMPA analyzer window was set to $w=1$, corresponding to a $3 \times 3$ pixel analysis region. A background region containing 2520 pixels was selected above the test object in the differential phase images captured with both detectors. This region was used to calculate the standard deviation, which was then converted into an angular sensitivity, as described in Sec.~\ref{sec:angular-sensitivity}. A comparison of angular sensitivities is provided Tab.~\ref{tab:detector_comparison}.

\begin{table}[htb]
    \centering
    \caption{Comparison of angular sensitivities (minimum resolvable refraction angle) of the two detector types for identical measurement conditions. Favorable values are marked in bold font.}
    \begin{tabular}{lcc}
        \hline
        & Photon counter & CMOS camera \\
        \hline
        Eff. pix. size [\si{\micro\meter}] & 3.83 & 3.66 \\
        $\sigma_x$ [\si{\nano\radian}] & \textbf{1.08} & 1.55 \\
        $\sigma_y$ [\si{\nano\radian}] & \textbf{1.11} & 1.49 \\
        $\sigma_{(x,y)}$ [\si{\nano\radian}] & \textbf{1.55} & 2.15 \\
        \hline
    \end{tabular}
    \label{tab:detector_comparison}
\end{table}

The results show that, given the same exposure time, the bidirectional angular sensitivity of the setup with the binned, fiber-coupled CMOS camera is $\SI{39}{\percent}$ lower than that of the setup with the photon-counting detector. This indicates a less efficient use of the incoming photons. Moreover, similar values for angular sensitivity in the $x$- and $y$-directions confirm that the stepping protocol resulted in bidirectional phase sensitivity. Our achieved angular sensitivity of \SI{1.55}{\nano\radian} compares favorably to the previously reported value of about \SI{143}{\nano\radian} in a parallel-beam configuration. However, the smaller effective pixel size of $\SI{0.96}{\micro\meter}$ needs to be considered~\cite{gustschin2021high}. We additionally note that high-resolution micro-computed tomography is typically performed with magnification lenses rather than fiber-coupling to reach smaller pixel sizes \cite{neldam2015application, longo2024syrmep}, which further reduces detector efficiency.

Although the reported sensitivity in the examined background region is high, it may be negatively affected inside the sample region due to Laplacian phase effects. Since UMPA only models first-order phase effects, i.e., changes in the refraction angle due to the sample, unaccounted-for second-order effects may lead to artifacts or noise. In this first demonstration of the setup, the effective propagation distance was close to the critical value of the near-field assumption ($z_\text{eff}/z_c \approx 0.5$).

\subsection{Tomographic measurement}
\begin{figure}
    \centering
    \includegraphics[width=1\linewidth]{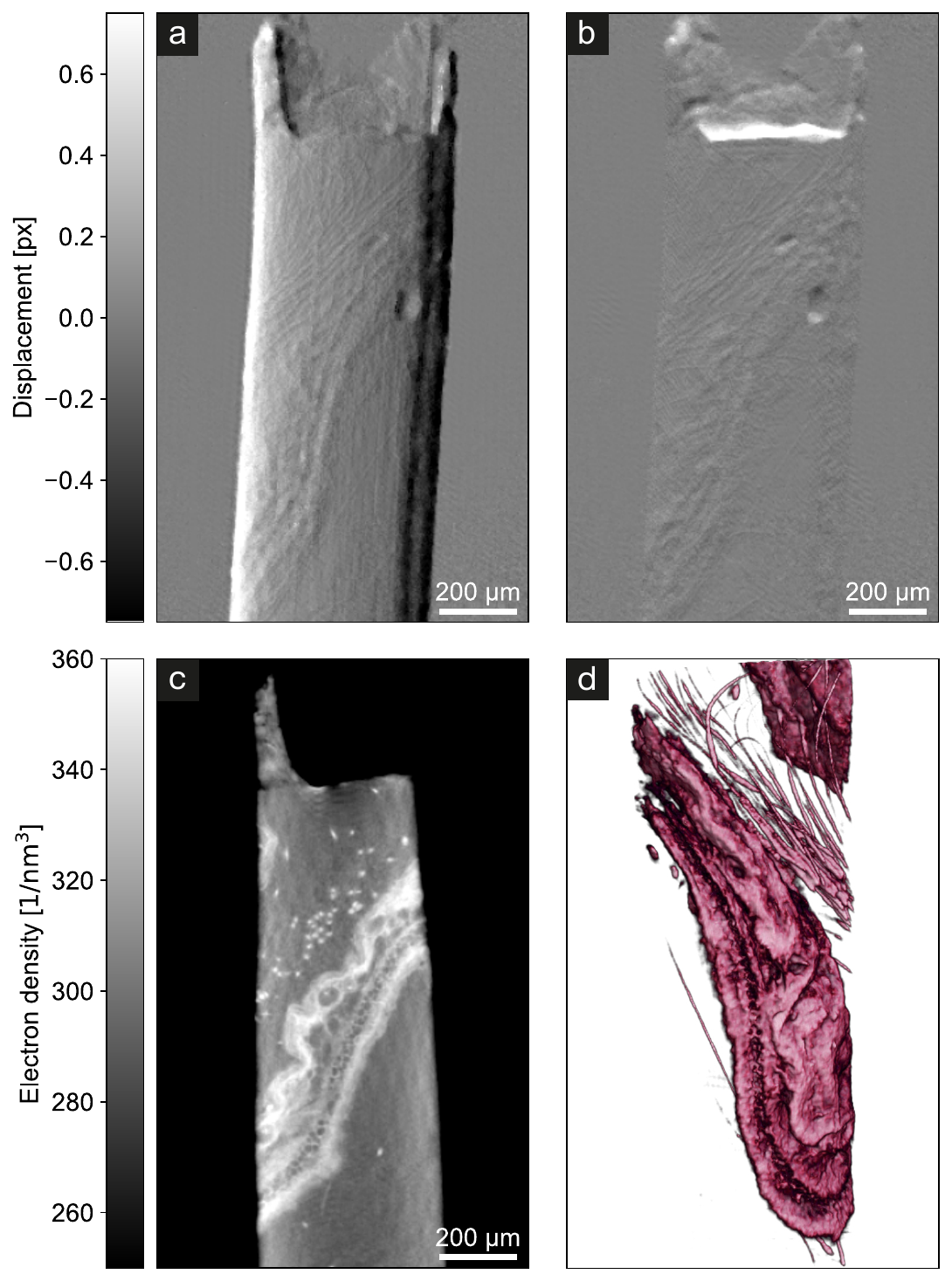}
    \caption{Tomographic phase scan of mouse skin embedded in a cylindrical piece of paraffin. (a) and (b) show projections of the differential phase signal projections in the $x$- and $y$-directions, respectively. These signals track the displacement of the reference pattern by the sample, in pixels, at the detector face. (c) shows a slice through the middle of the three-dimensional electron density map, obtained by 2D Fourier integration of the differential signals and tomographic reconstruction. (d) is a 3D rendering of the electron density volume.}
    \label{fig:mouse-scan}
\end{figure}

To demonstrate the capabilities of the combined setup for biological specimens, a piece of mouse skin embedded in a cylindrical block of paraffin wax was imaged. Two example differential phase projections are shown in Fig.~\ref{fig:mouse-scan}, panels (a) and (b). The differential phase signals in the $x-$ and $y-$directions illustrate the bidirectional phase sensitivity successfully achieved with the 1D stepping approach described in Sec.~\ref{sec:phase-stepping}. Panel (c) shows a slice through the reconstructed electron density volume. In the reconstruction, the different skin layers are well discernible against the paraffin. The 3D rendering in panel (d) shows the extent and position of the hairs. The electron density value of subcutaneous adipose tissue agrees well with previously reported literature values for white fat in the range of \SIrange{306}{312}{\per\nano\meter\cubed}~\cite{birnbacher2018electron}.

\section{Conclusion \& Outlook}
In this work, we have provided the first demonstration of modulation-based imaging in conjunction with X-ray waveguides, achieving visibility values of \SI{94.8}{\percent} for the CMOS and \SI{93.4}{\percent} for the photon-counting detector. This performance approaches the theoretical maximum of \SI{100}{\percent}, despite using modulators with small periods of at most \SI{10}{\micro\meter}. Our measurements reveal that high visibility is maintained over a broad range of effective propagation distances (approx.~\SIrange{100}{300}{\milli\meter}), which has useful practical implications for experimental flexibility.

The waveguide's magnification properties enable compatibility with large-pixel, photon-counting detectors while maintaining a spatial resolution similar to that of parallel-beam microtomography with conventional, indirect detectors. The combination of near-optimal quantum efficiency and high visibility directly addresses the dose problem in multimodal microtomography, opening new possibilities for quantitative, multimodal scans of samples in their native environments. Incorporating phase modulators provides access to the dark-field signal, which has not yet been explored in conjunction with X-ray waveguides. However, this signal was not included in our analysis, as no significant dark field is expected from our soft tissue sample. Future experiments should examine samples with small-angle scattering-inducing microstructure, possibly using either UMPA~\cite{zdora2017x} or a single-grid dark-field method for signal retrieval~\cite{how2022quantifying, croughan2023directional}.

Several aspects of the method can be optimized further. First, the flux reduction due to the waveguide can be mitigated by improving coupling with the KB mirror. However, next-generation synchrotron sources with their enhanced brilliance will naturally address this limitation. The current field of view can be expanded using waveguides with larger opening angles or by employing established stitching approaches~\cite{savatovic2024extending, john2024centimeter}. These methods will bring the technique closer to the capabilities of parallel-beam µCT. Additionally, the phase retrieval artifacts, which are likely caused by second-order phase effects, can be addressed by adapting the model to properly account for all wavefront changes induced by the sample.

Our results demonstrate that combining waveguide sources, Talbot array illuminators, and photon-counting detectors enables the simultaneous achievement of high spatial resolution and optimal dose efficiency. This work advances modulation-based imaging techniques and provides a platform to make the best use of fourth-generation synchrotron sources.

\section{Methods}
\subsection{Theoretical Considerations}
To utilize modulation-based imaging in conjunction with a waveguide focusing optic and a large-pixel photon-counting detector, several constraints must be considered. Since the algorithms tracking the sample-induced distortions of the phase modulator typically assume the X-ray near-field regime, the effective propagation distance may not become too large. A useful criterion for validating the near-field assumption is the critical propagation distance
\begin{equation}
    z_\text{c} = \frac{(2p_\text{eff})^2}{\lambda},
\end{equation}
where $p_\text{eff} = p / M$ is the effective detector pixel size (given the physical pixel size $p$ and the optical magnification $M$) and $\lambda$ is the X-ray wavelength~\cite{weitkamp2011ankaphase}. The propagation distance in the experiment shall be kept well below this value to minimize the presence of higher-order phase effects that are not accounted for in modulator-tracking algorithms.

Since the proposed setup makes use of a cone-beam geometry, the physical propagation distance between the sample and the detector $z$ needs to be replaced by an effective propagation distance $z_\text{eff}$. This follows from the Fresnel scaling theorem, which states that under the Fresnel approximation, an X-ray intensity field $I$ in a cone-beam geometry is related to an intensity field $I^{(P)}$ in a parallel-beam scenario via
\begin{equation}
    I(x,y,z = \Delta) = \frac{1}{M^2} I^{(P)}\left(\frac{x}{M}, \frac{y}{M}, z = \frac{\Delta}{M}\right).
\end{equation}
Here, $(x,y)$ are the transverse coordinates and $z = \Delta$ is the propagation distance~\cite{paganin2006coherent400}. The relevant quantity for calculating whether the setup is below the critical propagation distance is thus the effective propagation distance
\begin{equation}
    z_\text{eff} = \frac{\Delta}{M}.
\end{equation}

As an additional constraint, the magnification must be large enough to compensate for the larger pixel size of photon-counting detectors and enable imaging applications with effective pixel sizes in the common range for synchrotron-based µCT, i.e.,~\SIrange{1}{5}{\micro\meter}.
In summary, the constraints on the setup are (i) keeping the effective propagation distance below the critical value to stay in the near field and (ii) magnifying the sample enough to compensate for the larger pixel size of photon-counting detectors. In addition, (iii) optics with the highest possible opening angle should be used to enable larger fields of view. Theoretically, (iv) the chosen propagation distance should also coincide with a (fractional) Talbot distance $d_T$ of the phase modulator to maximize visibility, though it has been shown experimentally in Fig.~\ref{fig:visibility-plot} that this condition may be dropped.

\subsection{Visibility analysis}
Different approaches exist for calculating the visibility of a phase modulator~\cite{zdora2018state}. In this work, we make use of the visibility definition
\begin{equation}
    V = \frac{I_\text{max} - I_\text{min}}{I_\text{max} + I_\text{min}},
\end{equation}
where $I_\text{max}$ is the maximum and $I_\text{min}$ is the minimum recorded intensity in a subregion of the image, as previously used for structured modulators in Ref.~\cite{gustschin2021high}. We analyze the patch-wise visibility by subdividing the image into square regions whose area is large enough to capture a full modulator period. Since the effective pixel size changes in this geometry depending on the propagation distance, we use a patch size of $5 \times 5$ pixels for all measurements. This creates an analysis window large enough to capture at least one full modulator period at all examined effective pixel sizes. 

\subsection{Phase stepping}
\label{sec:phase-stepping}
For phase stepping, the modulator needs to be moved into different lateral positions. Bi-directional phase sensitivity requires that the intensity maxima created by the modulator are homogeneously distributed over the pixel matrix over the course of the phase stepping~\cite{gustschin2021high}. We make use of the unit cell sampling method discussed in Supplement 1 of Ref.~\cite{gustschin2021high}, which involves stepping the modulator in a diagonal fashion.

The tilt angle of this diagonal and the number of phase steps are calculated as follows~\cite{gustschin2021high}: Define two orthogonal vectors $(\textbf{a}, \textbf{b})$ in the modulator plane and choose their lengths in such a way that $a$ and $b$ are co-prime. The modulator is then moved along an axis oriented at an angle 
\begin{equation}
    \alpha = \arctan{(a/b)}
\end{equation}
to one direction of the grid pattern. When a number of $N = a^2 + b^2$ phase steps is performed, no redundant sampling occurs. The size of each individual phase step is defined by $d/N$, $d$ being the period of the modulator.

\subsection{Phase retrieval}
We employ the Unified Modulated Pattern Analysis (UMPA) algorithm to track sub-pixel displacements $(u_x, u_y)$ of the reference pattern caused by phase effects in the projection space~\cite{zdora2017x}. These local displacements of the pattern, which correspond to the refraction angles $(\alpha_x, \alpha_y)$, are converted from pixels on the detector face to a differential phase using
\begin{equation}
    \left(\frac{\partial\phi}{\partial x}, \frac{\partial\phi}{\partial y}\right) = \frac{2\pi}{\lambda} (\alpha_x, \alpha_y)= (u_x, u_y) \frac{2\pi}{\lambda} \, \frac{p_\text{eff}}{z_\text{eff}},
    \label{eq:pixels to refraction angles}
\end{equation}
where $\left(\frac{\partial\phi}{\partial x}, \frac{\partial\phi}{\partial y}\right)$ represent the $x$- and $y$-derivatives of the object's phase $\phi(x,y)$, $\lambda$ is the wavelength, and $p_\text{eff}$ is the effective pixel size~\cite{zdora2017x}. We then obtain the object's phase (up to an additive constant $C$) through 2D Fourier integration of the two differential phase signals~\cite{arnison2004linear, kottler2007two}:
\begin{equation}
    \phi(x, y) = \mathcal{F}^{-1}\left(\frac{\mathcal{F}(\frac{\partial\phi}{\partial x} + i \frac{\partial\phi}{\partial y})}{ik_x - k_y}\right) + C.
\end{equation}
Here, $\mathcal{F}$ denotes the Fourier transformation with respect to $x$ and $y$ while $\mathcal{F}^{-1}$ is the corresponding inverse Fourier transformation; $(k_x, k_y)$ are the Fourier-space coordinates associated with $(x,y)$. After defining the sample-free regions as having $\phi = 0$, the projected electron density is recovered using the relation
\begin{equation}
    \rho_{\text{e}, \perp}(x,y) = \frac{-\phi(x,y)}{\lambda\, r_e},
\end{equation}
where $r_\text{e}$ is the classical electron radius~\cite{paganin2006coherent114}. In a last step, the electron density distribution of the volume is obtained through tomographic reconstruction of the acquired projections. For reconstruction, we use the \emph{Core Imaging Library}~\cite{pasca_2024_12751447, jorgensen2021core} with the \emph{ASTRA} toolbox~\cite{van2016fast} backend.
To account for potential systematic measurement errors in UMPA, we calibrate the resulting electron density using the procedure described in Ref.~\cite{zandarco2024speckle}. We employ paraffin wax as a reference material to rescale the final electron density volume according to
\begin{equation}
\label{eq:calibration}
\rho_{e,\text{cal}}(x,y,z) = \frac{\rho_{e,\text{raw}}(x,y,z) - \bar{\rho}_{e,\text{bg}}}{\bar{\rho}_{e,\text{ref}} - \bar{\rho}_{e,\text{bg}}} \,\rho_{e,\text{theo}},
\end{equation}
where $\rho_{e,\text{cal}}(x,y,z)$ represents the calibrated three-dimensional electron density map, $\rho_{e,\text{raw}}(x,y,z)$ is the uncalibrated electron density map, $\bar{\rho}_{e,\text{bg}}$ and $\bar{\rho}_{e,\text{ref}}$ are the mean electron densities of the background and reference material, respectively, and $\rho_{e,\text{theo}}$ is the theoretical electron density of the reference material. Using a measurement at the P05 beamline calibrated with poly(methyl methacrylate) as the reference material, the electron density of paraffin (Paraplast Plus\textregistered) was determined to be \SI{299(7)}{\per\nano\meter\cubed}.

\subsection{Angular sensitivity}
\label{sec:angular-sensitivity}
The angular sensitivity of an X-ray imaging system is defined as the minimum detectable refraction angle $\alpha$ and is a widely used measure to compare different imaging systems~\cite{Modregger2011}. Considering Eq.~\eqref{eq:pixels to refraction angles}, it follows that
\begin{equation}
    (\sigma_x, \sigma_y) = \sqrt{\mathrm{Var}_\text{BG}\{(u_x, u_y)\}} \, \frac{p_\text{eff}}{z_\text{eff}},
    \label{eq:sensitivity}
\end{equation}
where $\mathrm{Var}_\text{BG}$ is the variance in a homogeneous region, typically the background without any sample in the beam~\cite[Supplement 1]{gustschin2021high}. Furthermore, the angular sensitivity is linked to other imaging quantities, such as visibility and the average number of counts, $N$. This relationship can be described by
\begin{equation}
    \sigma_{x/y} \propto \frac{1}{w \nu \sqrt{N}},
\end{equation}
where $v$ is the visibility and $w$ is the analyzer window size used by UMPA~\cite{Thuering_2012, zdora2017x}. 

\subsection{Corrections for beam stability}
Due to the strong intensity gradients induced by the modulators, even slight movements result in substantial changes in the recorded counts for each pixel. This is problematic because the phase-retrieval algorithm assumes that the beam profile is stable and all changes are related to sample properties~\cite{zdora2018state}.

To account for the movement, we decompose a number of consecutively recorded flat fields at each stepping position into eigenflats $E_i$ using principal component analysis, as described in Ref.~\cite{van2015dynamic}. Then, for each recorded projection image, we add the eigenflats using a weighted sum to best match the beam condition at the time of acquisition. This generates so-called dynamic flat-field images. The weights $w_i$ for this sum are obtained by comparing the weighted sum to a sample-free region $I_{\text{BG}}$ in each projection~\cite{riedel2023comparing}:
\begin{equation}
w_i = \underset{w_i}{\operatorname{argmin}} \left\| I_{\text{BG}} - \sum_{i} w_i\, E_i \right\|_2^2.
\end{equation}
Because a residual modulator pattern was still observed in the processed projections, the peaks corresponding to the modulator frequency in Fourier space were masked and set to zero prior to reconstructing the tomographic scan, thereby reducing these artifacts. However, this last step was not performed for the projections used for angular sensitivity calculations.

\subsection{Experimental Setup}
\label{sec:experimental-setup}
The experiment was conducted at the GINIX setup~\cite{kalbfleisch2010holography, kruger2012sub, doring2013sub} of the Coherence Beamline P10 (\emph{PETRA III}, DESY, Hamburg), which uses a U32 undulator as its X-ray source. The setup is depicted in Fig.~\ref{fig:setup}: After passing through a monochromator, the X-rays hit two total reflection mirrors arranged in a Kirkpatrick-Baez (KB) geometry within a vacuum vessel. The KB mirror provides a two-dimensional hard X-ray focus down to $250 \times$ \SI{250}{\nano\meter\squared}, which is used to focus the beam onto an X-ray waveguide. This creates a highly coherent source for the experiment. We used a waveguide channel with a width and depth of $d = \SI{100}{\nano\meter}$ and an optical depth of $L = \SI{1}{\milli\meter}$, defined by e-beam lithography in silicon (\emph{Eulitha GmbH}, Würenlos, Switzerland) and then capped by wafer bonding. The exit flux depends on storage ring operation, photon energy, beamline settings, and alignment. For the given experiment, it reached $\SI{1.5e9}{\per\second}$.

A holder mounted to three orthogonal linear stages was positioned at a distance $s$ downstream of the waveguide. This allowed for the insertion of different phase modulators and precise positioning both parallel to and perpendicular to the beam direction. The sample was placed on a rotation stage at a further distance $l$ downstream of the modulator.
Two detector systems were used to acquire images: The first was an Andor Zyla sCMOS camera (\emph{Oxford Instruments}, Tubney Woods, England) with a \SI{15}{\micro\meter} thick Gadox scintillator and magnifying optics. This resulted in an effective pixel size of \SI{6.5}{\micro\meter} for the array of $2160 \times 2560$ pixels. The second system used was an Eiger X4M photon-counting detector (\emph{DECTRIS AG}, Baden, Switzerland) with $2070 \times 2167$ pixels and a pitch of \SI{75}{\micro\meter}.
The phase modulators used were Talbot array illuminators fabricated from silicon, as described in Ref.~\cite{gustschin2021high}. Their design height imparts a $2\pi/3$ phase shift to portions of the wavefront at an energy of \SI{10}{\kilo\eV}. Two modulators with period sizes of \SI{7}{\micro\meter} and \SI{10}{\micro\meter} were employed. The first modulator creates a hexagonal lattice of foci, while the second creates a square lattice. For comparison, a piece of Kapton\textregistered\ B black was also used as a random diffuser.

For the tomography experiment, a mouse skin sample was placed at a distance $s = \SI{260}{\milli\meter}$ from the waveguide and imaged with the photon-counting detector. The beam energy was set to \SI{10}{\kilo\eV} and 250 different angles were captured over \SI{360}{\degree}, each with an exposure time of \SI{1.0}{\second}. After each full rotation of the sample, the modulator was moved to its next stepping position. A total of 50 reference images were acquired at each position. For this experiment, the modulator with a period of \SI{10}{\micro\meter} and square lattice pattern was used. To achieve bidirectional phase sensitivity with 1D linear phase stepping, the modulator was stepped along an axis inclined at an angle of \SI{11.3}{\degree} to the modulator pattern. This follows from selecting the two co-prime numbers $a = 1$ and $b = 5$, resulting in a total of $5^2 + 1^2 = 26$ modulator positions, as described in Sec.~\ref{sec:phase-stepping}. Phase retrieval was performed using a runtime-optimized version of UMPA~\cite{de2022high} with a window parameter of $w = 1$, corresponding to a window size of $3\times 3$ pixels. A projection to test phase sensitivity was acquired with the same modulator and settings, with the only differences being the number of reference images (10 instead of 50) and the sample, a silicon crystal with varying thicknesses.

\section{Funding}
The authors gratefully acknowledge financial support by the ERC Consolidator Grant (Julia Herzen, TUM, DEPICT, PE3, 101125761) and by the EIC Pathfinder (1MICRON, 101186826). The authors also acknowledge funding through the BMFTR joint project 05K25MG2, \emph{HoTop4: Optiken für die Ganzfeld-Bildgebung bei PETRA III und IV}.

\section{Acknowledgment}
We thank Benedikt Günther, Stefan Schwaiger, James Pollock, Jannis Ahlers, and Marie-Christine Zdora for fruitful discussions.

We acknowledge DESY (Hamburg, Germany), a member of the Helmholtz Association HGF, for the provision of experimental facilities under beamtime proposal I-20241169. This research was supported in part through the Maxwell computational resources operated at DESY.

\section{Disclosures}
The authors declare no conflicts of interest.

\section{Data availability} Data underlying the results presented in this paper are not publicly available at this time but may be obtained from the authors upon reasonable request.

\bibliography{bibliography}

\end{document}